\documentclass[aps,prd,preprint]{revtex4-1}
\usepackage{amssymb}
\usepackage{graphicx}
\usepackage[svgnames]{xcolor}
\usepackage[colorlinks=true,citecolor=blue,linkcolor=blue,citebordercolor=white,linkbordercolor=white]{hyperref}

\def\eq#1{{Eq.~(\ref{#1})}}

\def\fig#1{{Fig.~\ref{#1}}}

\begin{document}

\author{S.V. Novikov}
\affiliation{A.N. Frumkin Institute of
Physical Chemistry and Electrochemistry, Leninsky prosp. 31,
Moscow 119071, Russia}
\affiliation{National Research University Higher School of Economics, Myasnitskaya Ulitsa 20, Moscow 101000, Russia}

\title{Hopping charge transport in amorphous organic and inorganic materials with spatially correlated random energy landscape}

\begin{abstract}
General properties of the hopping transport of charge carriers in amorphous organic and inorganic materials are discussed. We consider the case where the random energy landscape in the materials is strongly spatially correlated. This is a very typical situation in the organic materials having the Gaussian density of states (DOS) and may be realized in some materials with the exponential DOS. We demonstrate that the different type of DOS leads to a very different functional form of the mobility field dependence even in the case of the identical correlation function of random energy. We provide important arguments in favor of the significant contribution of the local orientational order to the total magnitude of the energetic disorder in organic materials. A simple but promising model of charge transport in highly anisotropic composites materials is suggested.
\end{abstract}

\maketitle


\section{Introduction}

Last decades show a sharp surge in the number of papers devoted to various aspects of organic electronics. Interest to the area is mostly motivated and supported by the development of various devices: organic light emitting diodes, solar cells, field effect transistors, sensors, and others \cite{1,2,3,4,5}. For the efficient functioning of all such devices a fast transport of charge carriers (electrons and holes) is needed. Essentially, charge transport in amorphous molecular materials occurs as a series of elementary redox reactions (hops of the carrier) between two neighbor molecules. A very first hop from the conducting electrode to the adjacent molecule is a typical example of electrochemical reaction. Interplay between electrochemistry and charge carrier transport is important for the production of conducting organic materials. Many conducting polymers are obtained by the electrochemical oxidation of various organic compounds \cite{Hammerich-book,Kharton-book}. Kinetics of the growth of the conducting polymer film at the electrode is intimately related with the charge transport properties of the polymer \cite{Gelling-book}. In addition, our consideration of the transport properties of charge carriers in the medium with spatially correlated disorder may be applied  to the description of the ionic transport in amorphous solid state electrolytes \cite{Kuwata:417,Wang:1617,Knauth:911}.

Characteristics of a charge carrier transport in amorphous materials are defined by the statistics of the energetic and positional (or diagonal and non-diagonal) disorder. Energetic disorder describes random fluctuations of the energies $U(\vec{r})$ of the relevant levels of transport molecules, while the positional disorder describes fluctuations of the spatial positions of molecules. There is a general belief that for the quasi-equilibrium (or nondispersive\footnote{In the time-of-flight experiments charge carriers are initially generated in the vicinity of the electrode, then they drift under the action of the applied electric field to the opposite collecting electrode, and the transient electric current is observed. The term "dispersive" means that the current transient monotonously decreases with time revealing no characteristic features in the double linear current vs time coordinates. Oppositely, a nondispersive transport means that after the initial decay, associated with the energetic and spatial relaxation of the carriers, a well defined plateau of the current is formed indicating development of the quasi-stationary transport regime with the constant mean carrier velocity. Then the current demonstrates an abrupt decay related to the arrival of the majority of carriers to the collecting electrode. This is a fingerprint of the quasi-equilibrium charge transport. For the dispersive regime an energetic relaxation of the carriers is not finished and the transport is nonequilibrium \cite{disp}.}) charge transport the most important is the energetic disorder \cite{Nenashev:93201}. The reason is that the hopping rate $\Gamma\left(\vec{r}_{ij},\Delta U_{ij}\right)$ depends on distance $\vec{r}_{ij}=\vec{r}_i-\vec{r}_j$ between any given pair of transport molecules and the energy difference $\Delta U_{ij}=U_i-U_j$ between corresponding transport levels in a very different way. Typically,  $\Gamma$ is a symmetric function of $\vec{r}_{ij}$, $\Gamma(\vec{r}_{ij})=\Gamma(\vec{r}_{ji})$, and in the most cases \cite{Miller:745,Marcus:966}
\begin{equation}\label{dist}
\Gamma(\vec{r}_{ij}) \propto \exp\left(-2\frac{|\vec{r}_{ij}|}{r_0}\right)
\end{equation}
(here $r_0$ is a localization radius of the wave function of the transport molecule), while for the dependence $\Gamma(\Delta U_{ij})$ the microscopic balance takes place \cite{Chandler:1987book}, i.e.
\begin{equation}\label{micro}
\Gamma(\Delta U_{ij})=\Gamma(\Delta U_{ji})\exp\left(-\frac{\Delta U_{ij}}{kT}\right)
\end{equation}
and probability for a carrier to jump up in energy is restricted by the exponential factor in comparison to the probability of the downward hop. For the quasi-equilibrium transport the so-called "bad" sites give the most important contribution. "Bad" sites are sites which can keep carrier for the longest time (mean escape time is the longest). For the positional disorder the "bad" sites are sites well separated from the nearest neighbors, but the symmetry of $\Gamma$ means that the probability to get into such sites is pretty low. Hence, they are not very important. This is not the case for the energetic disorder, where "bad" sites are sites having energy much lower than the energies of the neighbor sites, the probability to get into those sites is not low, and the contribution of low energy sites to the typical transport time is significant. This is the reason for the primary importance of the energetic disorder.

The simplest and most important characteristic of the energetic disorder is the distribution density of random energies of transport levels, called the density of states. Typically, in amorphous organic materials the DOS has a Gaussian form \cite{Bassler:15,Pasveer:206601,Germs:165210}
\begin{equation}\label{Gauss}
P(U)=\frac{N_0}{(2\pi\sigma^2)^{1/2}}\exp\left(-\frac{U^2}{2\sigma^2}\right)
\end{equation}
with $\sigma\simeq$ 0.1 eV, while in inorganic materials it has an exponential form \cite{baranovski2006charge}
\begin{equation}\label{exp}
P(U)=\frac{N_0}{U_0}\exp\left(U/U_0\right),\hskip10pt U < 0,
\end{equation}
with $U_0\simeq$ 0.02 - 0.05 eV. Here $N_0$ is a total concentration of transport sites. Transport properties of the amorphous materials having Gaussian or exponential DOS are very different. The most striking difference is the eventual development of the quasi-equilibrium transport regime with constant average velocity $v$ and diffusivity $D$ for the Gaussian DOS for any temperature, while for the exponential DOS for low temperature $kT < U_0$ the process of the carrier energetic relaxation is infinite in time and average velocity decreases with time and, hence, with the thickness of the transport layer $L$, as $v\propto L^{1-U_0/kT}$ \cite{Rudenko:209}. Such dramatic difference in the transport behavior is explained by the very fact that for the Gaussian DOS the so-called density of occupied states $P_{\rm occ}(U)\propto P(U)\exp(-U/kT)$ does exists for any temperature, while for the exponential DOS this is not so for low temperature. The density $P_{\rm occ}(U)$ describes the distribution of the carriers in the quasi-equilibrium regime for $t\rightarrow\infty$. Moreover, using $P_{\rm occ}(U)$ we may immediately provide an estimation for the temperature dependence of the quasi-equilibrium drift mobility $\mu=v/E$ for the Gaussian DOS as $\ln\mu\propto -U_a/kT$, where $U_a=-U_{\rm max}=\sigma^2/kT$ is an effective activation energy of the carrier hops and $U_{\rm max}$ is the position of the maximum of $P_{\rm occ}(U)$. The resulting dependence $\ln\mu\propto -\left(\sigma/kT\right)^2$ is in a very good agreement with experimental data \cite{Bassler:15,Borsenberger:9,Schein:7295}.

Some time ago it was realized that the DOS is not the only characteristic of the random energy landscape which is relevant for the hopping charge transport. Possible spatial correlations in $U(\vec{r})$ are important as well. Correlations are mostly important for the electric field dependence of the carrier drift mobility $\mu(E)$. Electric field affects charge transport mostly by the variation of the $\Delta U_{ij}=\Delta U_{ij}^0-e\vec{E}\left(\vec{r}_i-\vec{r}_j\right)$ and this variation inevitable requires spatial displacement between sites $i$ and $j$. Typical magnitude of $\Delta U_{ij}^0$ strongly depends on the degree of correlation between $U_i$ and $U_j$ for a given distance $r_{ij}$, and that is why spatial correlation affects the dependence $\mu(E)$.

Correlation effects for the Gaussian DOS are studied well, especially for the case of 1D charge transport. Still, there are some problems that are not clearly understood and we are going to discuss them in the paper. Mostly, we are going to concentrate our attention on the verification of the important approximate analytical result of Deem and Chandler \cite{Deem:911} on the dependence of the transport parameters in the case of zero applied field on the dimensionality of space and discussion of the effect of the local orientational order in organic materials on the energetic disorder. For the exponential DOS correlation effects are not studied at all. Here we are going to provide a first consideration of the correlation effect on the mobility field dependence for the nondispersive quasi-equilibrium regime $U_0/kT < 1$. In addition, we suggest a simple model describing transport properties of the mesoscopically heterogeneous composite materials, where spatial correlations are naturally provided by the very structure of the material.

\section{Spatial correlation of the random energy landscape and charge transport in organic materials: Gaussian DOS}

Effect of the spatial correlation in the distribution of random energies $U(\vec{r})$ is best studied for the case of amorphous organic materials. In these materials it was found that the contribution from randomly located and oriented permanent dipoles and quadrupoles provides the Gaussian DOS with $\sigma\simeq 0.1$ eV and binary correlation function $C(\vec{r})=\left<U(\vec{r})U(0)\right>$ decaying as $1/r$ for the dipolar case and $1/r^3$ in the quadrupolar case, correspondingly \cite{Novikov:14573,Novikov:181,Novikov:89} (here angular brackets mean a statistical averaging). More precisely, contribution of dipoles and quadrupoles provides the Gaussian DOS only for the main body of the distribution and not so far tails, while for the far tails of the DOS its shape is not a Gaussian one \cite{Novikov:164510}. Deviation from the Gaussian shape is typically important only for the description of the charge carrier transport at very low (and, probably, experimentally inaccessible) temperature. Long range spatial correlation means that the amorphous organic material consists of large clusters where every cluster is a set of neighbor transport sites having close values of $U$ (visual representation of such medium and its comparison with the case of spatially noncorrelated Gaussian medium may be found in Ref. \cite{Novikov:41139}).

It was found that the correlation effectively governs the field dependence of the drift mobility. 1D transport model suggests that for the Gaussian DOS the power law correlation function $C(\vec{r})\propto 1/r^n$ leads to $\ln \mu \propto E^{n/(n+1)}$ \cite{Dunlap:542}. This relation for the case of dipolar and quadrupolar disorder has been confirmed by the extensive 3D computer simulation \cite{Novikov:4472,Novikov:104120,Novikov:89,Novikov:954}.

For some time it was a puzzle why the so-called Poole-Frenkel mobility field dependence $\ln \mu \propto E^{1/2}$ is ubiquitous in amorphous organic materials. For such dependence the exponent of the correlation function is $n=1$ which is perfectly suitable for polar organic materials with dipoles giving the major contribution to the total energetic disorder. In nonpolar materials we should expect that the major contribution is provided by quadrupoles and, correspondingly, $n=3$ giving $\ln \mu \propto E^{3/4}$. A possible solution of the puzzle is that for some reason all experimental studies of the mobility in nonpolar organic materials cover a limited field range, no more than one order of magnitude and typically even less \cite{Borsenberger:226,Borsenberger:555,Borsenberger:9,Sinicropi:331,Heun:245}. In such narrow field range it is not possible to make a reliable distinction between $\ln\mu\propto E^{1/2}$ and $\ln\mu\propto E^{3/4}$ dependences \cite{Novikov:181,Novikov:89,Novikov:2584}.

Majority of the fundamental results for the theory of multidimensional hopping charge carrier transport in amorphous materials has been obtained using computer simulations \cite{Bassler:15,Novikov:4472,Novikov:2584,Pasveer:206601}. Exact and nontrivial approximate analytical results are essentially limited to the consideration of 1D charge transport. In fact, we know only one important theoretical result which is valid for the case of multidimensional transport and obtained using the renormalization group (RG) method. Deem and Chandler \cite{Deem:911} showed that the leading asymptotics for the diffusivity in the zero applied field for the Gaussian DOS is
\begin{equation}\label{D(0)}
D(0)=D_0\exp\left[-\frac{1}{d}\left(\frac{\sigma}{kT}\right)^2\right],
\end{equation}
where $D_0$ is a diffusivity in the absence of the disorder, and $d$ is the dimensionality of the space. Due to the validity of the Einstein relation for $E=0$ even in the case of strong disorder, the similar relation is valid for the mobility
\begin{equation}\label{mu(0)}
\mu=\mu_0\exp\left[-\frac{1}{d}\left(\frac{\sigma}{kT}\right)^2\right].
\end{equation}

Relation (\ref{mu(0)}) provides two nontrivial statements: first, the particular  dependence of the exponent of the mobility (or diffusivity) on $d$ and, second,  independence of the diffusivity and mobility for $E=0$ of the correlation properties of $U(\vec{r})$. We can provide a limited verification of those statements. First of all, \eq{mu(0)} for $d=1$ is identical to the exact solution of 1D transport model \cite{Dunlap:542,Parris:2803}. For 3D case we can compare \eq{mu(0)} with the simulation data for three models having different correlation properties: short range correlation $C(\vec{r})=0$ for $r> 0$ (Gaussian Disorder Model (GDM) \cite{Bassler:15}), dipolar correlation $C(\vec{r})\propto 1/r$ (dipolar glass (DG) model \cite{Novikov:4472}), and quadrupolar correlation $C(\vec{r})\propto 1/r^3$ (quadrupolar glass (QG) model \cite{Novikov:949,Novikov:2532}). For these models
\begin{equation}\label{mu(0)-2}
\ln \mu/\mu_0 \propto -C\left(\frac{\sigma}{kT}\right)^2
\end{equation}
for $E\rightarrow 0$ and $C_{\rm GDM}\approx 0.38$, $C_{\rm DG}\approx 0.36$, and $C_{\rm QG}\approx 0.37$. We should note that these particular values of $C$ are valid only if we use the extrapolation of the mobility to $E=0$ according to the natural field dependence law, i.e.
\begin{equation}\label{extra}
\ln \mu/\mu_0 \approx
-C\left(\frac{\sigma}{kT}\right)^2+A E^m, \hskip10pt m=\frac{n}{n+1},
\end{equation}
where $n$ is the exponent in the dependence of the correlation function on distance $C(\vec{r})\propto 1/r^n$ (for the GDM formally $n=\infty$ and $m=1$) \cite{Novikov:2532}.
We see that values of $C$ for different models are indeed pretty close to the RG prediction $C=1/3$. Moreover, we should expect that this prediction gives only a leading term of the expansion of $C$ in powers of $1/d$, and there are corrections to that value proportional to higher powers of $1/d$. Part of the deviation between our values of $C$ and $1/d$ could be attributed to the very procedure of the calculation of $C$: it is obtained by the extrapolation of the dependence $\mu(E)$ to $E\rightarrow 0$ according to \eq{extra}, while the relation (\ref{mu(0)}) describes the exact value of $\mu$ at $E=0$. Hence, small difference between $1/3$ and actual values of $C$ is not surprising. We may conclude that our simulation data support the idea of the universality of the transport parameters in the limit $E\rightarrow 0$ irrelevant to the correlation properties of the Gaussian random energy landscape.

Another problem of the proper description of charge carrier transport in amorphous organic materials is related to the local orientational order in such materials. All models for the calculation of $\sigma$ (apart from the direct simulation of the microscopic structure of a particular amorphous organic material) treats the material as a regular lattice with sites randomly occupied (if the fraction of the occupied sites is less than 1) by randomly oriented organic molecules (in fact, very simple models of a molecule are used, such as point dipoles, quadrupoles, etc. \cite{Dieckmann:8136,Novikov:877e,Novikov:164510}) Moreover, usually it is assumed that the molecules are embedded in the continuous dielectric medium described by the only parameter, i.e dielectric constant $\varepsilon$. Such models neglect not only the local short range orientational correlations which are inevitable for large asymmetric organic molecules, typical for organic charge transport materials, but also an inapplicability of the macroscopic $\varepsilon$ for the description of the short range charge--dipole and charge--quadrupole interaction.

More reliable consideration of the dielectric properties of polar amorphous organic materials has been carried out by Madigan and Bulovi{\'c}  \cite{Madigan:216402}. They considered the lattice DG model and introduced a microscopic polarizability $\alpha$ to describe the dielectric properties of the medium. Hence, the local dipole moment depends on the local electric field
\begin{equation}\label{loc}
\vec{p}_i=\vec{p}_i^0+\alpha\vec{E}_i.
\end{equation}
Then the Claussius-Mossotti equation
\begin{equation}\label{CMeq}
\alpha=\frac{\varepsilon-1}{\varepsilon+2}\frac{3}{4\pi}V_m
\end{equation}
has been used to relate $\alpha$ and $\varepsilon$ (here $V_m$ is a volume per molecule). While for the simple DG model $\sigma\propto 1/\varepsilon$, Madigan and Bulovi{\'c}  found that for $\varepsilon\rightarrow\infty$ $\sigma$ goes to some nonzero constant value. The result is quite understandable because in this model the major contribution to $\sigma$ for large macroscopic $\varepsilon$ is provided by the short range interaction with neighbor dipoles effectively taken at $\varepsilon=1$.

\begin{figure}[tbp]
\begin{center}
\includegraphics[width=3in]{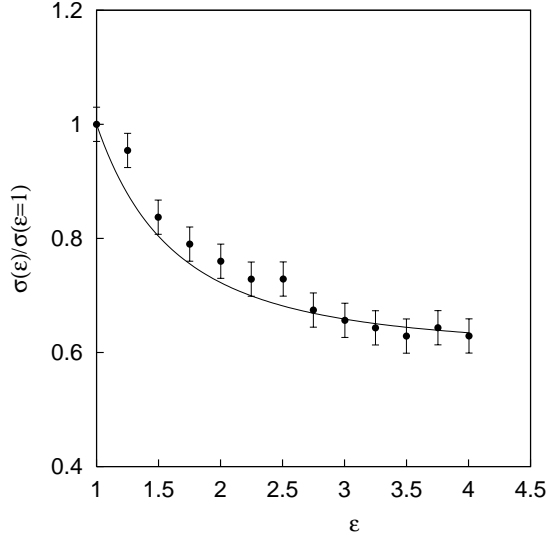}
\end{center}
\caption{Dependence of the ratio $\sigma(\varepsilon)/\sigma(\varepsilon=1)$ on $\varepsilon$ for the Madigan-Bulovi{\'c} model \cite{Madigan:216402} (dots). Solid line is calculated using \eq{s2_simple}.}
\label{CMmodel}
\end{figure}

It is interesting to note that the major result of Madigan and Bulovi{\'c} obtained by the time-consuming computer simulation (Fig. \ref{CMmodel}, dots) may be reproduced by a very simple calculation. For the lattice DG model with a single macroscopic $\varepsilon$ \cite{Novikov:877e}
\begin{equation}\label{s2}
 \sigma^2_d=\left<U^2(\vec{r})\right>=\frac{\sigma^2_0}{\varepsilon^2} S, \hskip10pt \sigma^2_0=\frac{e^2p^2c}{3a^4}, \hskip10pt S=\sum_{\vec{n}}\frac{1}{\left|\vec{n}\right|^4},
\end{equation}
where $a$ is the lattice scale, $p$ is the dipole moment, $c$ is the
fraction of sites occupied by dipoles, 3D vector $\vec{n}$ with
integer components runs over all lattice sites except the origin
$\vec{n}=0$, and for the simple cubic lattice (SCL) $S\approx 16.53$. If we assume that $\varepsilon=1$ for nearest dipoles with $|\vec{n}|=1$ and set the macroscopic $\varepsilon$ for all other sites, then
\begin{equation}\label{s2_simple}
 \frac{\sigma^2(\varepsilon)}{\sigma^2(\varepsilon=1)}=
  \left(1-\frac{N}{S}\right)\frac{1}{\varepsilon^2}+\frac{N}{S},
\end{equation}
where $N=6$ is the number of nearest sites for the SCL. This result is shown in Fig. \ref{CMmodel} (solid line). It is worth to note that for $\varepsilon=2-4$ (which is typical for organic materials) \eq{s2_simple} increases $\sigma(\varepsilon)$ by the factor of 2 in comparison to \eq{s2}. Hence, according to that calculation, typical magnitude of the disorder in amorphous organic materials should be $\simeq 0.2$ eV. This increase is almost entirely provided by the contribution of nearest neighbors, and that contribution is a very short range correlated one in comparison to the long range correlation described by the dipolar correlation function $C(\vec{r})\approx 0.76 \sigma^2_d a/r$, $r\gg a$ \cite{Novikov:14573}. This means that for the calculation of the correlation function for $r\gg a$ we have to use $\sigma_d$, estimated by the old \eq{s2}, while the total $\sigma_{\rm tot}$ (magnitude of the total disorder in the DG model) approximately obeys \eq{s2_simple}.

Computer simulation for the 3D case suggests that for the DG model the mobility dependence on $T$ and $E$ has the form
\begin{equation}\label{mu}
\ln \mu/\mu_0\approx -\left(\frac{3\sigma_{\rm
tot}}{5kT}\right)^2+C_E\left[\left(\frac{\sigma_d}{kT}\right)^{3/2}-\Gamma\right]\sqrt{eaE/\sigma_d},
\end{equation}
where $C_E\approx 0.78$, $\Gamma\approx 2$, and the mobility temperature dependence at low fields is governed by the total disorder $\sigma_{\rm tot}$, while the mobility field dependence at moderate fields is governed by the correlated component of the disorder, i.e. by $\sigma_d$ \cite{Novikov:4472,Novikov:2584}. Hence, taking into account the Madigan-Bulovi{\'c} correction, we should have the same mobility field dependence but much stronger mobility temperature dependence. Experimental data do not support this suggestion. Typically, the value of $\sigma$, estimated from the temperature dependence of $\mu$ for low fields is not significantly greater that $\sigma$ estimated from the field dependence of $\mu$, and  $\sigma_{\rm tot}$ is still close to 0.1 eV \cite{Bassler:15,Borsenberger:9,Novikov:4472,Novikov:2584}. The most natural explanation of the discrepancy is a contribution from the short range local order in amorphous organic materials which reduces $\sigma_{\rm tot}$ but gives no significant correction to $\sigma_d$. We believe that the results of the paper \cite{Madigan:216402} clearly indicate an importance of local short range order for the development of the random energy landscape in organic materials.

\begin{figure}[tbp]
\begin{center}
\includegraphics[width=3in]{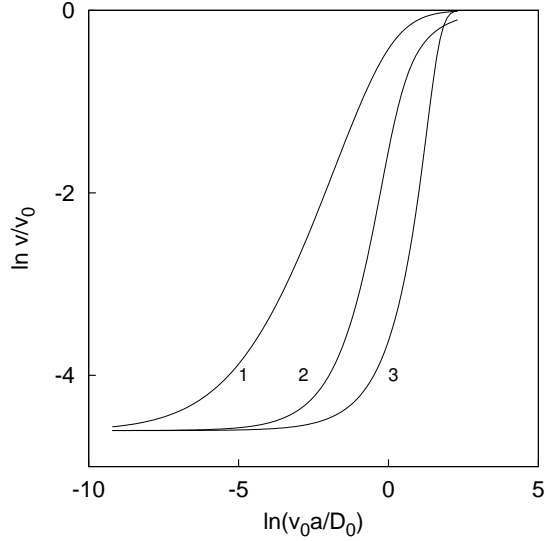}
\end{center}
\caption{Mobility field dependence \eq{v-inf} for the spatially correlated exponential DOS and various kinds of the correlation function $c^2(x)$: $a^2/(x^2+a^2)$ (curve 1), $\exp(-x/a)$ (curve 2), and $\theta(a-x)$ (curve 3), correspondingly; for all curves $1-K^2=1\times 10^{-2}$. Note that $v_0 a/D_0=eaE/kT$.} \label{mixt}
\end{figure}

\section{Spatial correlation of the random energy landscape and charge transport in inorganic materials: exponential DOS}

Amorphous inorganic materials are very different from the organic materials. For such materials the exponential DOS described by \eq{exp} is ubiquitous.
To the best of our knowledge an effect of the correlated random energy landscape has not been considered for the exponential DOS. The obvious difficulty is a problem of introducing the correlation in the exponential distribution. To overcome this difficulty we use a trick borrowed from the probability theory, i.e. we a going to produce the correlated exponential distribution using the auxiliary Gaussian ones \cite{Devroye-book}. Indeed, if $X$ and $Y$ are two independent and identically distributed random Gaussian variables with zero mean and unit variance, then
\begin{equation}
U=-\frac{U_0}{2}\left(X^2+Y^2\right)
\label{U-XY}
\end{equation}
has the exponential distribution (\ref{exp}). If the Gaussian variables $X$ and $Y$ are correlated ones, i.e. $\left<X(x)X(0)\right>=c_X(x) \neq 0$, here $x$ is a spatial variable, then $U(x)$ has a correlated distribution with the correlation function
\begin{equation}\label{2a-corrU}
c_U(x)=\left<U_1 U_2\right>-\left<U\right>^2=U_0^2\left[c_X^2(x)+c_Y^2(x)\right],
\end{equation}
and it is obvious that in this way we can model any positive binary correlation function for the random field $U(x)$.

\begin{figure}[tbp]
\begin{center}
\includegraphics[width=3in]{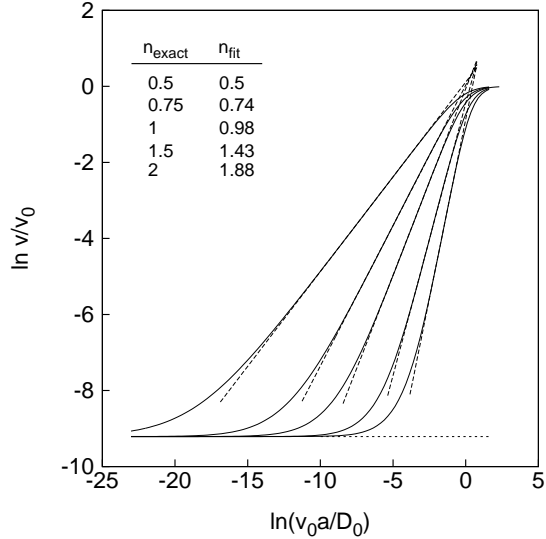}
\end{center}
\caption{Mobility field dependence (solid lines) calculated by \eq{v-inf} for the power law correlation function $c^2(x)=a^n/(x^2+a^2)^{n/2}$ for various $n$: 0.5, 0.75, 1, 1.5, and 2, from the upmost curve to the bottom, correspondingly, and for $1-K^2=1\times 10^{-4}$. Dotted line shows the limiting dependence $v/v_0=1-K^2$, and the broken lines show fits for power law dependence $v/v_0\propto v_0^{n}$.} \label{XY_fig1}
\end{figure}

A very good approximation for the long time behavior of the hopping transport is a model of carrier diffusion in the random energy landscape $U(x)$. In 1D case the stationary solution of the diffusion equation gives for the average carrier velocity \cite{Parris:5295}
\begin{equation}\label{v}
v=\frac{D_0\left(1-e^{-\gamma L}\right)}{\int\limits_0^L dx \hskip2pt\exp\left(-\gamma x\right)Z(x,L)}, \hskip10pt Z(x,L)=\frac{1}{L}\int\limits_0^L dy \hskip2pt\exp\left[\frac{U(y)-U(x+y)}{kT}\right], \hskip10pt \gamma=v_0/D_0.
\end{equation}
Here $v_0$ and $D_0$ are carrier velocity and diffusivity in the absence of the disorder, and $L$ is a thickness of the transport layer. We can immediately calculate the average carrier velocity in the limit $L\rightarrow\infty$, because in that limit
\begin{equation}\label{Z-inf}
Z(x)=\lim_{L\rightarrow\infty} Z(x,L)=\left<\exp\left[\frac{U(0)-U(x)}{kT}\right]\right>
\end{equation}
and
\begin{equation}\label{v-inf}
v=\frac{D_0}{\int\limits_0^\infty dx \hskip2pt\exp\left(-\gamma x\right)Z(x)}.
\end{equation}
Using the probability distribution for the correlated Gaussian variables
\begin{equation}\label{Gauss}
P_G(X_1,X_2)=\frac{1}{2\pi\sqrt{1-c^2}} \exp\left(-\frac{X_1^2+X_2^2-2cX_1X_2}{2(1-c^2)}\right),\hskip10pt \left<X_1X_2\right>=c,
\end{equation}
with the same relation for $Y$ (we assume for simplicity $c_X=c_Y=c$) and taking into account relation (\ref{U-XY}), we obtain
\begin{equation}
Z(x)=\frac{1}{1-K^2\left[1-c^2(x)\right]}, \hskip10pt K=U_0/kT.
\label{Z)}
\end{equation}
For $x\rightarrow \infty$ $c(x)\rightarrow 0$, while $c(0)=1$. Hence, for $\gamma\rightarrow\infty$ $v\rightarrow v_0$, while for $\gamma\rightarrow 0$ $v\rightarrow v_0(1-K^2)$. We immediately see that for the infinite medium $v$ could be nonzero only for $K <1$, and for $K=1$ there is a transition to the dispersive non-equilibrium regime. This result also means that the reliable determination of the functional form of the dependence $v(v_0)$ can be carried out only in the close vicinity to the transition to the dispersive regime where variation of the ratio $v/v_0$ is significant. Field dependence of the dimensionless mobility $v/v_0=\mu/\mu_0$ for various kinds of the correlation function $c^2(x)$ is shown in Fig. \ref{mixt}.

We can calculate the field dependence of the average velocity (actually, the dependence of $v$ on the bare velocity $v_0$ which is proportional to $E$) for the power-law correlation function $c^2(x)= a^n/(x^2+a^2)^{n/2}$ (this particular form is inspired by the correlation properties of the amorphous organic materials). Using a saddle point method we obtain an intermediate asymptotics
\begin{equation}
v\simeq \frac{v_0}{\sqrt{2\pi n}}\left(\frac{\gamma a e}{n}\right)^n, \hskip10pt v_0(1-K^2) \ll v \ll v_0,
\label{2a-Sp3}
\end{equation}
which agrees well with the direct calculation using \eq{v-inf} (see Fig. \ref{XY_fig1}). We see that the asymptotics (\ref{2a-Sp3}) is developing only for $1-K^2 \ll 1$.

Calculation of the carrier drift velocity for the dispersive non-equilibrium regime $K > 1$ is a much more difficult task. It will be considered in a separate paper.

At the moment we have no direct experimental evidence for the existence of the spatial correlations in amorphous materials having the exponential DOS. Quite probably, this situation is partly explained by the lack of the enthusiasm to search for the correlations, because previously nobody emphasized their importance and possible effect on the mobility field dependence. Nonetheless, in the recent paper \cite{May:136401} it was demonstrated that there is a possibility to find the exponential DOS in some organic amorphous materials, and such materials typically have strongly correlated random energy landscape. We may expect that the correlated exponential DOS may be found not in inorganic, but rather in organic materials.

\begin{figure}[ht]
\begin{center}
\includegraphics[width=3in]{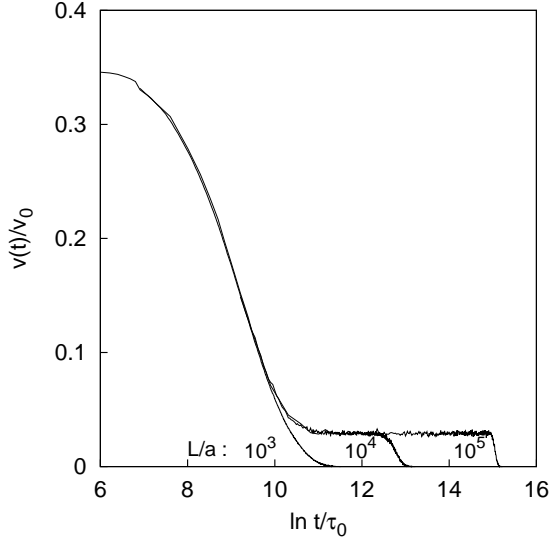}
\end{center}
\caption{Photocurrent transients in the composite material for the bulk regime  and various thickness $L$ of the transport layer (shown near the corresponding curve). Other parameters are $l_a/a=100$, $l_b/a=200$, $\tau_a/\tau_0=100$, $\tau_b/\tau_0=1$, $v_a/v_0=0.01$, and $v_b/v_0=1$. Here $a$ is a spatial scale (e.g., size of a molecule), $\tau_0$ is a time scale, and $v_0=a/\tau_0$. For the thick layers one can see formation of the quasi-equilibrium transport regime having a velocity plateau.}
\label{sim1}
\end{figure}

\begin{figure}[ht]
\begin{center}
\includegraphics[width=3in]{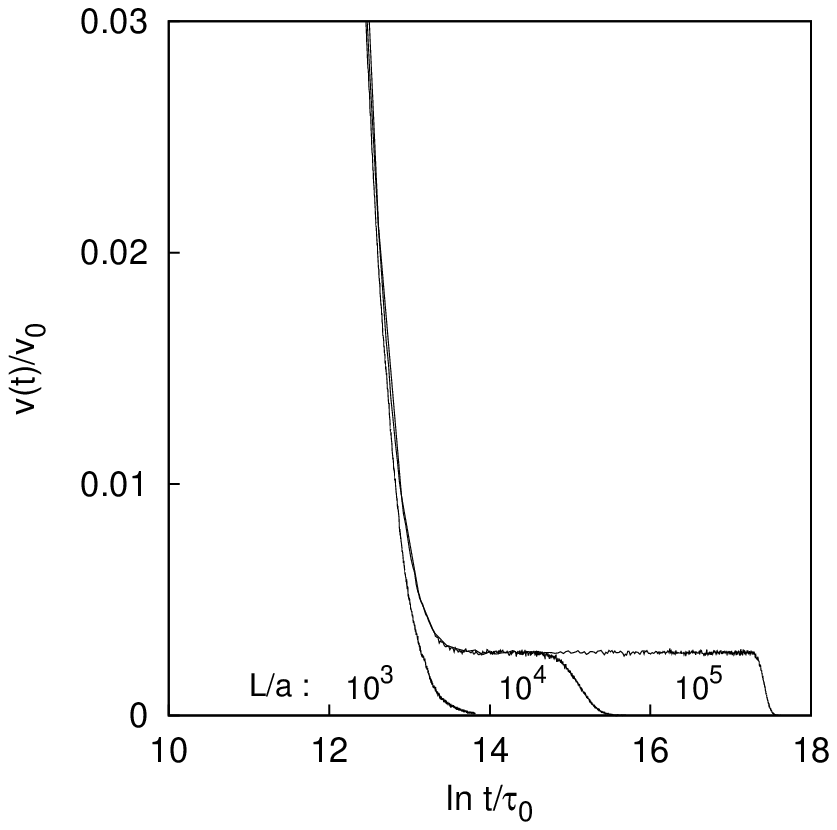}
\end{center}
\caption{Photocurrent transients in the composite material for the interface regime and various thickness $L$ of the transport layer (shown near the corresponding curve). Here $\tau_a/\tau_0=1\times 10^5$ and all other parameters are the same as in \fig{sim1}. Again,  for the thick layers we see formation of the quasi-equilibrium transport regime having a velocity plateau.}
\label{sim2}
\end{figure}

\section{Charge transport in composite organic materials}

A very different type of spatial correlation may arise in another class of transport materials, i.e. composite materials. Composite materials are mesoscopically inhomogeneous materials, typically having large domains with very different properties. If we assume that every domain may be characterized by its own transport properties, then we have a transport medium with spatially correlated distribution of transport parameters, such as carrier velocity, diffusivity, etc.

Let us consider a simplest model of highly anisotropic composite material having chain-like structure with all chains elongated in the same direction, the hopping charge transport occurs along the chains and is essentially one-dimensional. Electric field is oriented parallel to the chains and every chain is composed by the clusters of two types of transport materials, material A and material B. We assume that there are distributions of the clusters on length, $p_a(l)$ and $p_b(l)$, and the the carrier motion in the clusters is a pure drift, characterized by velocities $v_a$ and $v_b$. In addition, we assume that there are distributions of time $p_a(\tau)$ and $p_b(\tau)$ to cross the interfaces between clusters (index \textit{a} here means that the carrier goes from cluster A to cluster B, and index \textit{b} means the transition B $\rightarrow$ A).

We can write a simple formula for the average carrier velocity for the infinite medium
\begin{equation}\label{v_inf}
v_\infty=\frac{\left<l_a\right>+\left<l_b\right>}
{\frac{\left<l_a\right>}{v_a}+\frac{\left<l_b\right>}{v_b}+\left<\tau_a\right>+\left<\tau_b\right>}.
\end{equation}
The structure of the denominator indicates that there are two very distinct transport regimes. In the bulk regime first and second terms in the denominator dominate (most relevant is the time for a carrier to drift over clusters), while in the interface regime third and forth terms are more important (time to overcome interfaces is dominating). For the diffusivity the corresponding result is
\begin{equation}\label{D_inf}
D_\infty=\frac{v_\infty^2\left<(\delta t)^2\right>}{2\left<t\right>}=v_\infty^2
\frac{\frac{\left<l_a^2\right>-\left<l_a\right>^2}{v_a^2}+\frac{\left<l_b^2\right>-\left<l_b\right>^2}{v_b^2}+\left<\tau_a^2\right>-\left<\tau_a\right>^2+\left<\tau_b^2\right>-\left<\tau_b\right>^2}
{2\left({\frac{\left<l_a\right>}{v_a}+\frac{\left<l_b\right>}{v_b}+\left<\tau_a\right>+\left<\tau_b\right>}\right)},
\end{equation}
here $t$ is the time needed to a carrier to travel across the sample having some finite but very large thickness $L$ and $\delta t$ is the fluctuation of that time (more precisely, we have to consider the limit $L\rightarrow\infty$). Again, we can differentiate between the bulk and interface regimes, but in addition we have a mixed regime. Indeed, we may quite easily imagine a situation where the fluctuation of the clusters' lengths is negligible $\left<l_{a,b}^2\right>-\left<l_{a,b}\right>^2\rightarrow 0$, while at the same time the dominant contribution to the denominator in \eq{D_inf} still comes from the bulk terms. The opposite situation $\left<\tau_{a,b}^2\right>-\left<\tau_{a,b}\right>^2\rightarrow 0$ is rather improbable: typical fluctuation of $\tau_{a,b}$ could be considered originating from the energetic barriers between clusters, in such a case $p(\tau)\propto\exp(-\tau/\tau_0)$ with $\tau_0\propto \exp(-\Delta/kT)$, where $\Delta$ is the height of the barrier. For the exponential distribution $\left<\tau^2\right>-\left<\tau\right>^2=\left<\tau\right>^2$ and variance of $\tau$ can be negligible only if $\left<\tau\right>$ is negligible.

\begin{figure}[ht]
\begin{center}
\includegraphics[width=3in]{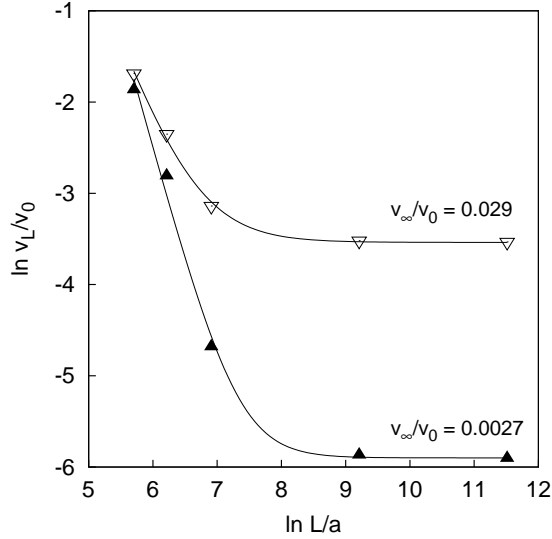}
\end{center}
\caption{Dependence of the average carrier velocity $v_L$ in the composite material on the thickness $L$ of the transport layer for the bulk ($\triangledown$) and interface ($\blacktriangle$) regime. Solid curves show the best fit for \eq{v_L}.}
\label{sim3}
\end{figure}

Some transport characteristics may be very similar for both regimes (e.g., general shape of current transients), yet there is no necessity to concentrate our attention exclusively on the transport properties, so we can expect many possible significant differences. For example, suppose that there is some  chemical reaction between charge carriers and molecules of the medium. In the bulk regime carrier spends most time in the interior regions of the clusters, while for the interface regime it dwells mostly near the interfaces. Properties of the material could be very different in those domains and we may expect different kinetic regimes. Moreover, many purely transport characteristics should be very different in the bulk and interface regimes. Indeed, in the bulk regime, especially for the case of large clusters, the dependence of $v$ on the applied electric field $E$ is expected to have the Poole-Frenkel form $\ln (v/E) \propto E^{1/2}$ because velocities $v_a$ and $v_b$ obey this very law. In the interface regime we should expect the dependence $\ln (v/E) \propto E$ because the reduction of the barrier height is proportional to $E$.

Transport properties for the finite thickness $L$, especially the shapes of photocurrent transients, are impossible to calculate analytically. For this reason we carried out computer simulation and some typical results are shown in Figs. \ref{sim1} and \ref{sim2}. With the increase of $L$ the average carrier velocity $v_L$ decreases and its behavior agrees well with the relation
\begin{equation}\label{v_L}
v_L=v_\infty\left[1+\left(\frac{L_0}{L}\right)^n\right]
\end{equation}
(see \fig{sim3}), while the limit velocity $v_\infty$ perfectly agrees with \eq{v_inf}.

Evidently, this simple model could demonstrate a wide variety of transport properties. A very tempting task should be to search for the transition between quasi-equilibrium and non-equilibrium transport regimes, resembling the transition between nondispersive and dispersive regimes. Quite probably, such transition could occur only for a very specific kind of the probability distributions $p_a(\tau)$ and $p_b(\tau)$ (for example, having long tails or exponentially wide relevant domain of $\tau$).

\section{Conclusion}

We considered effects of a spatial correlation of the random energy landscape in amorphous materials on the transport properties of such materials. Statistical properties of the energy landscape are of crucial importance for the hopping charge carrier transport. In organic materials the DOS usually has the Gaussian shape, while in inorganic materials it usually has the exponential shape. Correlations mostly affect the mobility field dependence. We demonstrated that the resulting mobility field dependences are principally different for the Gaussian and exponential DOS even for the same type of the binary correlation function of the random energy. We demonstrated that the experimental search of the effects of correlation for the exponential DOS in the nondispersive regime should be carried out in a very close vicinity to the transition to the dispersive regime. We argued that the local orientational order significantly reduces the total energetic disorder in polar organic materials. We also suggested the simple model for the description of transport properties of highly anisotropic composite materials. In spite of its simplicity, the model demonstrate very rich behavior and is promising for further development.

\section*{Acknowledgement}

Partial financial support from the RFBR grant 15-53-45099-IND-a is acknowledged.



\end{document}